\documentclass[english,a4paper,oneside,11pt]{article}
\usepackage[T1]{fontenc}
\usepackage[latin9]{inputenc}
\usepackage{amsmath}
\usepackage{babel}

\usepackage{amsmath,amstext,amsfonts,amsbsy,amssymb,amscd,bbm,epsfig,lscape}

\begin{document}

\begin{titlepage}

\vspace*{-45truemm}
\begin{flushright}
\hspace*{-1cm} RM3-TH/14-5, ROM2F/2014/02 
 
\end{flushright}\vspace{9truemm}


\centerline{\Large \bf  $D^0-\overline{D}^0$ Mixing in the Standard Model and Beyond } 
\centerline{\Large \bf from $\mathbf{N_f=2}$ Twisted Mass QCD}

\vskip 9 true mm
\centerline{ N.~Carrasco$^{(a)}$, M.~Ciuchini$^{(a)}$, P.~Dimopoulos$^{(c,d)}$, R.~Frezzotti$^{(c,e)}$, }
\centerline{ V.~Gimenez$^{(f)}$, V.~Lubicz$^{(a,b)}$, G.C.~Rossi$^{(c,e)}$, F.~Sanfilippo$^{(h)}$, } 
\centerline{ L.~Silvestrini$^{(g)}$, S.~Simula$^{(a)}$, C.~Tarantino$^{(a,b)}$ }

\vspace*{1truemm}
\begin{figure}[!h]
  \begin{center}
    \includegraphics[scale=0.70]{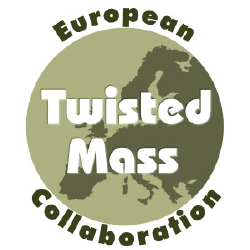}
 \end{center}
\end{figure}
\vspace*{2truemm}

\vskip 1 true mm
\vskip 0 true mm

\centerline{\it $^{(a)}$ INFN, Sezione di Roma Tre}
\centerline{\it Via della Vasca Navale 84, I-00146 Rome, Italy}
\vskip 2 true mm
\centerline{\it $^{(b)}$ Dipartimento di Matematica e Fisica, Universit\`a  Roma Tre}
\centerline{\it Via della Vasca Navale 84, I-00146 Rome, Italy}
\vskip 2 true mm
\centerline{\it $^{(c)}$ Dipartimento di Fisica, Universit\`a di Roma ``Tor Vergata''}
\centerline{\it Via della Ricerca Scientifica 1, I-00133 Rome, Italy}
\vskip 2 true mm
\centerline{\it $^{(d)}$ Centro Fermi - Museo Storico della Fisica e Centro Studi e Ricerche Enrico Fermi}
\centerline{\it Compendio del Viminale, Piazza del Viminale 1 I-00184, Rome, Italy}
\vskip 2 true mm
\centerline{\it $^{(e)}$ INFN, Sezione di ``Tor Vergata"}
\centerline{\it Via della Ricerca Scientifica 1, I-00133 Rome, Italy}
\vskip 2 true mm
\centerline{\it $^{(f)}$  Departament de F\'{\i}sica Te\`orica and IFIC, Univ. de Val\`encia-CSIC}
\centerline{\it Dr.~Moliner 50, E-46100 Val\`encia, Spain}
\vskip 2 true mm
\centerline{\it $^{(g)}$ INFN, Sezione di Roma, Piazzale A. Moro, I-00185 Rome, Italy}
\vskip 2 true mm
\centerline{\it $^{(h)}$ School of Physics and Astronomy, University of Southampton}
\centerline{\it SO17 1BJ Southampton, United Kingdom}
\vskip 2 true mm

\vskip 15 true mm


\begin{abstract}{
We present the first unquenched lattice QCD results for the bag parameters controlling the short 
distance contribution to $D$ meson oscillations in the Standard Model and beyond. 
We have used the gauge configurations produced by the European Twisted Mass Collaboration with $N_f=2$ dynamical quarks, at four lattice spacings and light meson masses in the range 
$280 \div 500$ MeV. Renormalization is carried out 
non-perturbatively with the RI-MOM method. The bag-parameter results have been used 
to constrain New Physics  effects in $D^0-\bar D^0$ mixing, to put a lower bound
to the generic New Physics scale and to constrain off-diagonal squark mass terms for TeV-scale
Supersymmetry.
}
\end{abstract}
\end{titlepage}

\section{Introduction}
\label{sec:intro}

The study of meson oscillations currently represents one of the most powerful probes in searching for 
New Physics (NP).
The $K$ and $B_{(s)}$ systems are well studied experimentally and all the 
available data are compatible with the Standard Model (SM) predictions.
Improved theoretical predictions and future experiments will be important to look for possible NP effects with higher accuracy. 
The phenomenon of $D^0 - \bar D^0$ mixing has been established only in 
2007~\cite{Aubert:2007wf,Staric:2007dt}.
As it involves mesons with up-type quarks, it is complementary to $K$ and 
$B_{(s)}$ oscillations in providing information on NP.
From the theory side, $D^0 - \bar D^0$ mixing has the disadvantage of being affected by large long-distance effects, related to the down and strange quarks circulating in the box diagrams.
Only order of magnitude estimates exist for the long-distance contributions and 
they are at the level of the experimental constraints. However, the SM contribution
to $D^0-\bar D^0$ mixing is real to very high accuracy.
Therefore, in spite of the SM uncertainty, significant constraints on NP can be obtained in this sector
from CP-violating observables~\cite{Gabbiani:1996hi,
Raz:2002ms,Ciuchini:2007cw,Nir:2007ac,Blanke:2007ee,He:2007iu,Golowich:2007ka,Bona:2007vi,
Dutta:2007ue,Hiller:2008sv,Chen:2009xjb,Gedalia:2009kh,Kagan:2009gb,Altmannshofer:2009ne,
Altmannshofer:2010ad,Isidori:2010kg,Crivellin:2010ys,Guadagnoli:2010sd,Calibbi:2012yj,
Bevan:2012waa}.
These constraints rely on the lattice computation of the bag-parameters of four-fermion operators describing
$D^0 - \bar D^0$ mixing beyond the SM.

In this paper we use the $N_f=2$ gauge
configurations~\cite{Baron:2009wt,Blossier:2010cr}, generated by the European 
Twisted Mass Collaboration (ETMC), at four values of the lattice spacing to obtain continuum limit estimates for 
the full basis of $\Delta C=2$ four-fermion operators. 
This is the first unquenched calculation of the whole set of $D$ meson bag-parameters.

The most general $\Delta C=2$ effective Hamiltonian of dimension-six operators is

\begin{equation}
H_{\textrm{eff}}^{\Delta C=2}=\frac{1}{4}{\displaystyle \sum_{i=1}^{5}C_{i}(\mu)Q_{i}}(\mu),
\label{eq:Heff}
\end{equation}
where $\mu$ is the renormalization scale and $C_i$ are the model-dependent 
Wilson coefficients encoding the short distance contributions. The operators 
$Q_i$ involving light ($\ell$) and charm ($c$) quarks read, in the so-called 
SUSY basis,
\begin{equation}
\begin{array}{l}
Q_{1}=\left[\bar{c}^{a}\gamma_{\mu}(1-\gamma_{5})\ell^{a}\right]\left[\bar{c}^{b
}\gamma_{\mu}(1-\gamma_{5})\ell^{b}\right]\,,\\
Q_{2}=\left[\bar{c}^{a}(1-\gamma_{5})\ell^{a}\right]\left[\bar{c}^{b}(1-\gamma_{
5})\ell^{b}\right]\,, \\
Q_{3}=\left[\bar{c}^{a}(1-\gamma_{5})\ell^{b}\right]\left[\bar{c}^{b}(1-\gamma_{
5})\ell^{a}\right]\,,\\
Q_{4}=\left[\bar{c}^{a}(1-\gamma_{5})\ell^{a}\right]\left[\bar{c}^{b}(1+\gamma_{
5})\ell^{b}\right]\,, \\
Q_{5}=\left[\bar{c}^{a}(1-\gamma_{5})\ell^{b}\right]\left[\bar{c}^{b}(1+\gamma_{
5})\ell^{a}\right]\,,
\label{eq:Qi}
\end{array} 
\end{equation}
where $a$, $b$ are color indices and Dirac indices (understood) are contracted within brackets.
In the SM only $Q_1$ enters the effective Hamiltonian.

According to the Operator Product Expansion (OPE), the long-distance non-perturbative QCD
contributions are enclosed in the matrix elements of the renormalized four-fermion operators, 
which can be written in terms of bag-parameters as
\begin{equation}
\begin{array}{l}
\langle\overline{D}^{0}|{Q}_{1}(\mu)|D^{0}\rangle=\xi_{1}{B}_{1}(\mu)m_{
D}^{2}f_{D}^{2}\,,\\
\langle\overline{D}^{0}|{Q}_{i}(\mu)|D^{0}\rangle=\xi_{i}{B}_{i}
(\mu)\left[\dfrac{m_{D}^{2}f_{D}}{{\mu}_{c}(\mu)+{\mu}_{\ell}(\mu)}
\right]^{2}\,,\qquad \textrm{for}\quad i\,=\, 2, \dots, 5\,,
\end{array}
\label{eq:xi}
\end{equation} 
where $\xi_i=\{8/3,-5/3,1/3,2,2/3\}$.

For the reader's convenience we give in Table~\ref{tab:BD-results} our final results for the ${B}_{i}$ bag-parameters, quoting the total uncertainty 
(statistical and systematic added in quadrature). 
From these results one notices moderate deviations from the vacuum insertion approximation (the size of which is, of course, scheme dependent) which are much smaller than in the kaon system but larger than for $B$-mesons~\cite{Carrasco:2013jda}.
\begin{table}[!t]
\begin{centering}
\begin{tabular}{||c||c|c|c|c|c|}
\hline 
 & $B_{1}$ & $B_{2}$ & $B_{3}$ & $B_{4}$ & $B_{5}$\tabularnewline
\hline 
\hline 
$\overline{\rm{MS}}$ (3GeV) & 0.75(02) & 0.66(02) & 0.96(05) & 0.91(04) & 1.10(05)\tabularnewline
\hline 
RI-MOM (3GeV) & 0.74(02) & 0.82(03) & 1.21(06) & 1.09(05) & 1.35(06)\tabularnewline
\hline 
\end{tabular}
\par\end{centering}
\caption{\label{tab:BD-results} {\it Results for the bag-parameters of 
$\overline{D}^{0}-D^{0}$
mixing, renormalized in the $\overline{\rm{MS}}$ scheme of ref.~\cite{mu:4ferm-nlo} and in the RI-MOM scheme at 3 GeV.}}
\end{table}

As in our recent works on $K$ and $B_{(s)}$ 
mixing~\cite{Bertone:2012cu,Carrasco:2013zta}, we use the results obtained for the full 
set of $\Delta C=2$ bag-parameters to improve the bounds on the NP scale
coming from $D$-meson mixing, following the method of 
ref.~\cite{Bona:2007vi}. We also recompute the bounds on off-diagonal squark masses
from gluino-mediated contributions to $D^0-\bar D^0$ mixing in the Minimal Supersymmetric
Standard Model (MSSM), updating the analysis presented in ref.~\cite{Ciuchini:2007cw}.
As for the experimental results we use the recent average of $D$-meson mixing data 
computed by the UTfit collaboration~\cite{Bevan:2014tha}.

The plan of the paper is as follows.
In Section~\ref{sec:pheno}, based on the results of this work for the $\Delta 
C=2$ bag-parameters, we discuss the bounds coming from $D$-meson mixing on the NP scale
and on off-diagonal squark mass terms.
In Section~\ref{sec:lat_simul} we give details about the lattice simulation and we describe the techniques that have been used in this 
work. 
In Section~\ref{sec:results} we discuss the continuum and chiral extrapolation 
and we present the results for the bag-parameters of the full four-fermion operator basis. 
We collect in Appendix~\ref{sec:table_results} the lattice bare bag-parameters for all the quark mass combinations and $\beta$ values we had available.  

\section{Bounds on the NP scale and on the squark mass terms}
\label{sec:pheno}

$\Delta F=2$ processes provide some of the most stringent constraints
on NP generalizations of the SM. Several phenomenological analyses of
$\Delta F=2$ processes have been performed in the last years, both for
specific models and in model-independent
frameworks~\cite{Bona:2007vi,Ligeti:2010ia,Buras:2010pz,Lenz:2010gu,
  Lunghi:2010gv,Adachi:2011cb,Calibbi:2012at,KerenZur:2012fr,Mescia:2012fg,Buras:2012dp,
  Bertone:2012cu}. While the SM prediction for $B_{(s)}^0-\bar
B_{(s)}^0$ mixing and $\varepsilon_K$ is theoretically well under control, the SM contribution to $D^0-\bar D^0$ mixing is
plagued by long-distance contributions. However, due to the SM flavor
structure, CP violation in $D^0-\bar D^0$ mixing receives negligible
SM contributions. Therefore, significant constraints on NP can be obtained in this sector
from CP-violating observables.

In two previous papers~\cite{Bertone:2012cu,Carrasco:2013zta} we have
presented the first unquenched ($N_f=2$) lattice QCD results in the continuum
limit for the matrix elements of the operators describing $K$ and
$B_{(s)}$ oscillations in extensions of the SM.  In the same papers we
have updated the generalization of the Unitarity Triangle analysis including possible NP effects,
improving the bounds coming from $K^0-\bar K^0$ and $B_{(s)}^0-\bar
B_{(s)}^0$ mixings.

In a similar way, we present here the first unquenched ($N_f=2$) lattice QCD
results for the bag-parameters of the full $\Delta C =2$
four-fermion operators basis and we use them to improve the bounds coming
from $D$-meson mixing on the NP scale and on the off-diagonal squark
mass terms, updating the analysis in
refs.~\cite{Ciuchini:2007cw,Bona:2007vi}. As for the experimental
results, we use the recent averages of $D$-meson mixing data derived by
the UTfit collaboration~\cite{Bevan:2014tha}. With the latest experimental
updates, the imaginary part of the $D$ mixing amplitude is very
strongly constrained, leading to very tight bounds on possible
CP-violating NP contributions to the mixing, as shown in Table~\ref{tab:D}. 


Let us first discuss the model-independent analysis. The most general
effective weak Hamiltonian for $D$ mixing of dimension six
operators is parameterized by Wilson coefficients of
the form
\begin{equation}
  C_i (\Lambda) = \frac{F_i L_i}{\Lambda^2}\, ,\qquad i=1,\ldots,5\, ,
  \label{eq:cgenstruct}
\end{equation}
where $F_i$ is the (generally complex) relevant NP flavor coupling,
$L_i$ is a (loop) factor which depends on the interactions that
generate $C_i(\Lambda)$, and $\Lambda$ is the NP scale, i.e.\ the
typical mass of new particles mediating $\Delta C=2$ transitions. For
a generic strongly interacting theory with an unconstrained flavor
structure, one expects $F_i \sim L_i \sim 1$, so that the
phenomenologically allowed range for each of the Wilson coefficients
can be immediately translated into a lower bound on
$\Lambda$. Specific assumptions on the flavor structure of NP
correspond to special choices of the $F_i$ functions.

Following ref.~\cite{Bona:2007vi}, in deriving the lower bounds on
the NP scale $\Lambda$, we assume $L_i = 1$, that corresponds to
strongly-interacting and/or tree-level coupled NP. Two other
interesting possibilities are given by loop-mediated NP contributions
proportional to either $\alpha_s^2$ or $\alpha_W^2$. The first case
corresponds for example to gluino exchange in the MSSM. The second case applies to all models with SM-like
loop-mediated weak interactions. To obtain the lower bound on
$\Lambda$ entailed by loop-mediated contributions, one simply has to
multiply the bounds we quote in the following by
$\alpha_s(\Lambda)\sim 0.1$ or $\alpha_W \sim 0.03$.

\begin{table}[!h]
\begin{center}
\begin{tabular}{|@{}ccc|}
\hline
 & $95\%$ upper limit  &
Lower limit on $\Lambda$ \\
&(GeV$^{-2}$) &
 (TeV)\\
\hline \hline
\phantom{A}Im\,$C^D_1$ & $[-0.9,2.5] \cdot 10^{-14}$ & $6.3 \cdot 10^{3}$ \\
\phantom{A}Im\,$C^D_2$ & $[-2.8,1.0] \cdot 10^{-15}$ & $1.9 \cdot 10^{4}$  \\
\phantom{A}Im\,$C^D_3$ & $[-3.0,8.6] \cdot 10^{-14}$ & $3.4 \cdot 10^{3}$  \\
\phantom{A}Im\,$C^D_4$ & $[-2.7,8.0] \cdot 10^{-16}$ & $3.5 \cdot 10^{4}$  \\
\phantom{A}Im\,$C^D_5$ & $[-0.4,1.1] \cdot 10^{-14}$ & $9.5 \cdot 10^{3}$ \\
\hline
\end{tabular}
\end{center}
\caption {{\it $95\%$ probability intervals for the imaginary part of the Wilson coefficients, Im\,$C^D_i$,
  and the corresponding lower bounds on the NP scale, $\Lambda$, for
  a generic strongly interacting NP with generic flavor structure 
($L_i=F_i=1)$.}} 
\label{tab:D}
\end{table}
\begin{figure}[!h]
\begin{center}
\includegraphics[bb=171bp 497bp 496bp 705bp,scale=0.6]{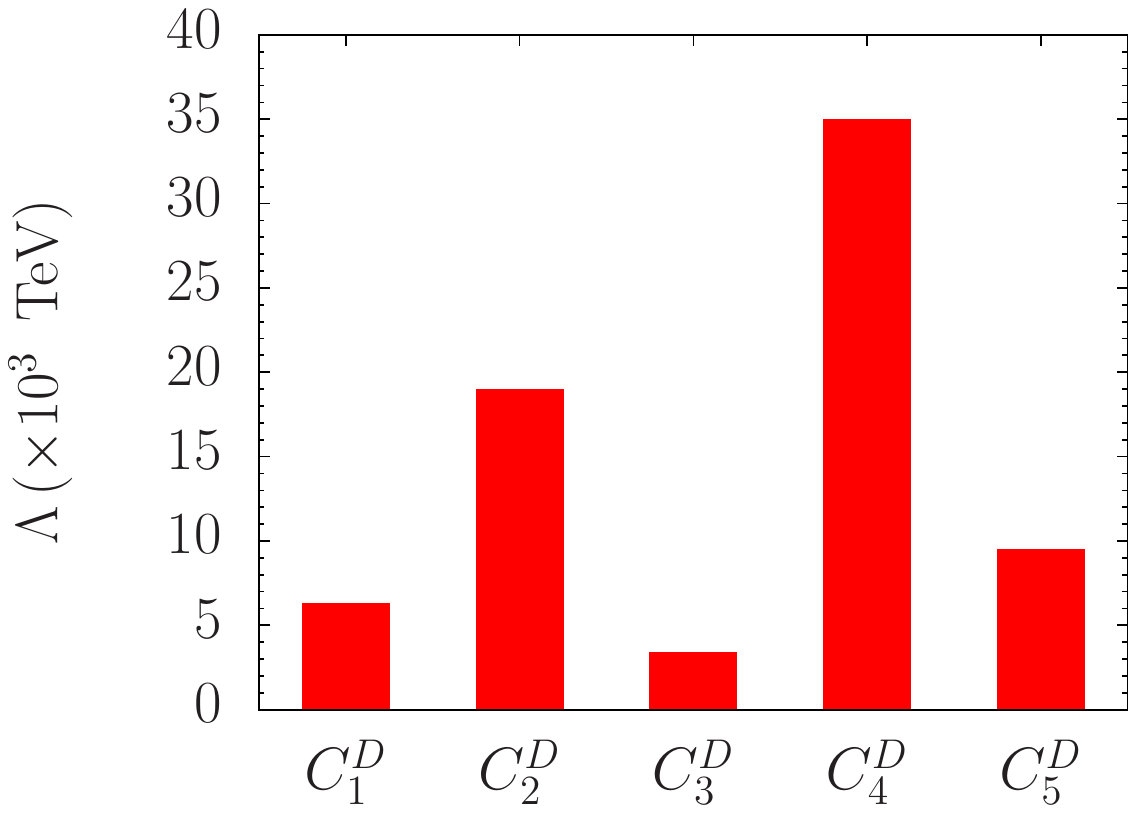} 
\end{center}
\caption{\label{fig:bounds} {\it Lower bounds on the NP scale as obtained from the constraints on the imaginary part of the Wilson coefficients, Im\,$C^D_i$.}}
\end{figure}
The results for the upper bounds on the imaginary part of the Wilson coefficients, Im\,$C^{D}_i$, and
the corresponding lower bounds on the NP scale $\Lambda$ are collected
in Table~\ref{tab:D}. The latter are also shown in fig.~\ref{fig:bounds}. The superscript $D$ is to recall that we
are reporting the bounds coming from the $D$-meson sector we are here
analyzing.

We remind the reader that the analysis is performed (as
in ref.~\cite{Bona:2007vi}) by switching on one coefficient at the time in
each sector, thus barring the possibility of accidental
cancellations among the contributions of different operators.
Therefore, the reader should keep in mind that the bounds may be
weakened if, instead, some accidental cancellation occurs.

In comparison with the analyses in refs.~\cite{Bertone:2012cu,Carrasco:2013zta}, we confirm that the most stringent constraints on the NP scale
 come from the $K^0-\bar{K}^0$ matrix elements, while the bounds coming from  $D^0-\bar{D}^0$ are more stringent than those coming from $B^0-\bar{B}^0$. 
 
We now turn to Supersymmetry (SUSY), and consider a general MSSM with arbitrary
off-diagonal squark mass terms. In this framework the dominant
contribution to flavor changing neutral current (FCNC) processes is expected to come from gluino
exchange, since the quark-squark-gluino vertex is proportional to
$g_s$ and involves both chiralities, generating all the operators in
eq.~(\ref{eq:Heff}). Therefore, we study the constraints on the
off-diagonal mass terms connecting up- and charm-type squarks of
helicities $A$ and $B$ in the super-CKM basis, normalized to the
average squark mass, denoted by $ \left( \delta^u_{12} \right)_{AB}$.
The bounds scale linearly with the average squark mass, up to
logarithmic terms due to QCD evolution. For reference, we report
the constraints obtained for gluino and average squark masses of
$1$ TeV. As above, we only quote the constraints obtained from the
CP-violating part of the $\Delta C=2$ amplitude, which correspond to
bounds on the imaginary part of $ \left( \delta^u_{12}
\right)_{AB}^2$. A constraint on the real part could be obtained by
making an educated guess on the size of the SM contribution; however,
we prefer to stick to model-independent results in the present
analysis. 

\begin{table}[!t]
\begin{center}
\begin{tabular}{|ccc|}
\hline
$\sqrt{
  \left\vert \mathrm{Im}
\left(
  \delta_{12}^u \right)_{LL,RR}^2\right\vert}$ & $\sqrt{
  \left\vert \mathrm{Im}
\left(
  \delta_{12}^u \right)_{LR,RL}^2\right\vert}$ & $\sqrt{
  \left\vert \mathrm{Im}
\left(
  \delta_{12}^u \right)_{LL=RR}^2\right\vert}$ \\
\hline \hline
0.019 & 0.0025 & 0.0011\\
\hline
\end{tabular}
\end{center}
\caption{{\it Upper bounds at $95\%$ probability on $\sqrt{
    \left\vert \mathrm{Im} \left(\delta_{12}^u \right)_{AB}^2
    \right\vert}$ for squark and gluino masses equal to 1 TeV. The three bounds are respectively obtained assuming: i) a dominant LL (or RR) mass insertion, ii) a dominant LR (or RL) mass insertion, iii) $ \left( \delta^u_{12} \right)_{LL}=\left( \delta^u_{12} \right)_{RR}$.}}
\label{tab:SUSY}
\end{table}
We use the mass-insertion approximation for degenerate squarks at the
NLO in QCD~\cite{rm1:4ferm-nlo,mu:4ferm-nlo,Ciuchini:2006dw} (see
ref.~\cite{Virto:2009wm} for the results of the SUSY matching in the
mass-eigenstate basis). The bounds are reported in Table
\ref{tab:SUSY} (see
refs.~\cite{Gabbiani:1996hi,Ciuchini:2007cw,Nir:2007ac,Golowich:2007ka,Hiller:2008sv,Gedalia:2009kh,Altmannshofer:2009ne,Altmannshofer:2010ad,Crivellin:2010ys,Calibbi:2012yj}
for previous analyses). Since there is no SM contribution, the bounds
on the SUSY $ \left( \delta^u_{12} \right)_{AB}$ are invariant under the exchange of chiralities.

We cannot compare directly the present bounds in Table~\ref{tab:SUSY} with
our previous results~\cite{Ciuchini:2007cw} which reported bounds on the absolute
values of the $ \left( \delta^u_{12} \right)_{AB}$ using an estimate of the long-distance contributions.
For the sake of comparison, we have checked that following the same procedure
as in ref.~\cite{Ciuchini:2007cw} we obtain bounds stronger by a factor from 3
to 5.

\section{Lattice setup and simulation details}
\label{sec:lat_simul}
The $N_f=2$ gauge configuration ensembles employed in the present analysis have been generated by the ETM Collaboration. 
The four values of the simulated lattice spacing lie in the interval [0.05, 0.1] fm. 
Dynamical quark simulations have been performed using the tree-level improved Symanzik gauge action~\cite{Weisz:1982zw}
and the Wilson twisted mass action~\cite{Frezzotti:2000nk} tuned to maximal twist~\cite{FrezzoRoss1}.
More details on the action and our $N_f=2$ gauge ensembles can be found in refs.~\cite{Boucaud:2007uk, Boucaud:2008xu, Baron:2009wt,Blossier:2010cr}. 
We stress that the use of maximally twisted fermionic action offers the advantage  of automatic O$(a)$  improvement for  
all the interesting physical observables computed on the lattice~\cite{FrezzoRoss1}. 

For the evaluation of the four-fermion matrix elements on the lattice we use a mixed fermionic action setup
where we adopt different regularizations for sea and valence quarks as  proposed in Ref.~\cite{Frezzotti:2004wz}.
This particular setup offers the advantage that one can compute matrix elements that are at the same time 
O$(a)$-improved and free of wrong chirality mixing effects~\cite{Bochicchio:1985xa}. 
These two properties have already proved to be very beneficial in the
study of neutral $K$- and $B$-meson oscillations~\cite{Bertone:2012cu,Constantinou:2010qv,Dimopoulos:2009es,Carrasco:2011gr,Carrasco:2013zta,Carrasco:2013jaa}. 

We have computed 2- and 3-point correlation functions with valence quark masses ranging from the light sea quark mass up to around the physical charm quark mass. 
Simulation details are given in Table~\ref{tab:runs}, where  $\mu_{\ell}$ and $\mu_c$  indicate  the 
bare light and charm-like valence quark masses respectively. 
The values of the light valence quark mass are set equal to the light sea ones, $a\mu_{\ell} = a\mu_{sea}$, and they 
correspond to light pseudoscalar mesons in the range $280 \div 500$~MeV. 
\begin{table}[!t]
\begin{centering}
\begin{tabular}{cccc}
\hline 
{\small $\beta$} & {\small $a^{-4}(L^{3}\times T)$} & {\small $a\mu_{\ell}=a\mu_{\textrm{sea}}$} & {\small $a\mu_{c}$}  \tabularnewline
\hline 
{\small 3.80} &{\small $24^{3}\times48$} & {\small 0.0080, 0.0110} & {\small 0.1982, 0.2331, 0.2742} \tabularnewline
{\small $a\sim0.098\mbox{fm}$ } &  &  &   \tabularnewline
\hline 
{\small 3.90} & {\small $24^{3}\times48$} & {\small 0.0040, 0.0064, 0.0085, 0.0100} & {\small 0.1828, 0.2150, 0.2529} \tabularnewline
{\small $a\sim0.085\mbox{fm}$} &   {\small $32^{3}\times64$} & {\small 0.0030, 0.0040} &  \tabularnewline
\hline 
{\small 4.05} &{\small $32^{3}\times64$} & {\small 0.0030, 0.0060, 0.0080} & {\small 0.1572, 0.1849, 0.2175}  \tabularnewline
{\small $a\sim0.067\mbox{fm}$} &  &  &      \tabularnewline
\hline 
{\small 4.20}  & {\small $32^{3}\times64$} & {\small 0.0065} & {\small 0.13315, 0.1566, 0.1842} \tabularnewline
{\small $a\sim0.054$fm} &   {\small $48^{3}\times96$} & {\small 0.0020} &  \tabularnewline
\hline 
\end{tabular}
\end{centering}
\centering{}\caption{\label{tab:runs} {\it Simulation details for correlator computation  at four values of the inverse gauge coupling $\beta = 3.80, 3.90, 4.05 
~\rm{and}~ 4.20$. The quantities $a\mu_{\ell}$ and $a\mu_c$ stand for  
light and charm-like bare valence quark mass values respectively, expressed in lattice units.}}
\end{table}

Renormalised quark masses are obtained from the bare ones
using the renormalisation constant $Z_{\mu} = Z_P^{-1}$~\cite{Frezzotti:2000nk,Frezzotti:2004wz}, whose values have been computed in~\cite{Constantinou:2010gr,Carrasco:2013zta} using RI-MOM 
techniques. 
The physical values for the light and charm quark mass are $\bar{m}_{u/d}(2\,\rm{GeV})=3.6(2)$ MeV and  $\bar{m}_{c}(m_c)=1.28(4)$ GeV, taken from ref.~\cite{Blossier:2010cr}.

We have computed 2- and 3- point correlation  functions by employing smearing techniques on a set of 100-240 independent gauge 
configurations for each ensemble and evaluated statistical errors using a bootstrap method\footnote{The bootstrap method also serves the purpose of taking into account correlations over different timeslices.}.
Smeared interpolating operators become
mandatory in the presence of relativistic heavy (charm-like and heavier) quarks. 
Smearing turns out to reduce the coupling of the
interpolating field with the excited states, thus increasing its projection 
onto the lowest energy eigenstate. The usual drawback, {\it i.e.} the increase of
the gauge noise due to fluctuations of the links entering in the smeared
fields, is controlled by replacing thin gauge links with APE smeared ones.   
With this technical improvement heavy-light meson masses
and matrix elements can be extracted at relatively small temporal separations while keeping
noise-to-signal ratio under control. We employed  
Gaussian smearing~\cite{Gusken:1989qx, Jansen:2008si} for heavy-light meson
interpolating fields at the source and/or the sink. The smeared
field is of the form:
\begin{equation}
\Phi^{\rm{S}}  = (1+6\kappa_{{\rm G}})^{-N_{\rm{G}}} 
 (1 + \kappa_{\rm{G}} a^2 \nabla^2_{\rm{APE}})^{N_G} \Phi^{\rm{L}},
\end{equation}
where $\Phi^{\rm{L}}$ is a standard local source and $\nabla_{\rm{APE}}$ 
is the lattice covariant derivative with APE 
smeared gauge links characterised by the parameters  $\alpha_{\rm{APE}}=0.5$ and
$N_{\rm{APE}}=20$. We have taken $\kappa_{\rm{G}}=4$ and $N_{\rm{G}}=30$. 
We have noticed that in practice a better signal to noise ratio is found 
when the source, rather than the sink,
is smeared. Thus 2-point Smeared-Local (SL) correlation functions yield better improved 
plateaux for the lowest energy mass state than Local-Smeared (LS) or Smeared-Smeared (SS) ones.
In a recent paper~\cite{Carrasco:2013zta}  ETMC investigated optimised 
interpolating  operators for both three- and two-point correlation functions both of which enter 
in the computation of the bag parameters. 
It has been found out 
that within the statistical uncertainty (which is at the level of $1\%$ or less) no difference can be seen between the optimised 
and the simple smeared interpolating fileds. 
In the present work we use the same lattice data as those in ref.~\cite{Carrasco:2013zta}.
For this reason we are confident that excited states are well suppressed for our plateau choices.

\begin{figure}[!t]
\begin{center}
\begin{tabular}{cc}
\hspace*{0.3cm}\vspace*{0.35cm}\includegraphics[bb=171bp 497bp 496bp 705bp,scale=0.6]{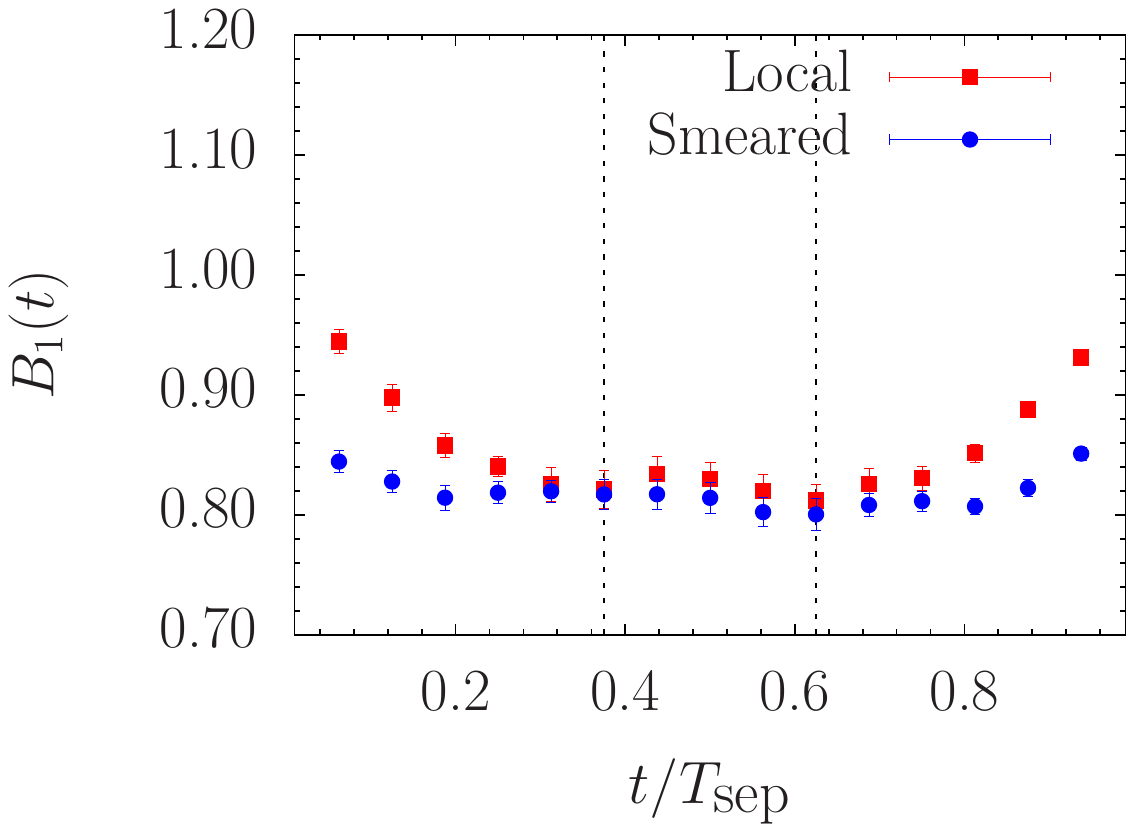} & \vspace*{0.35cm}\includegraphics[bb=171bp 497bp 496bp 705bp,scale=0.6]{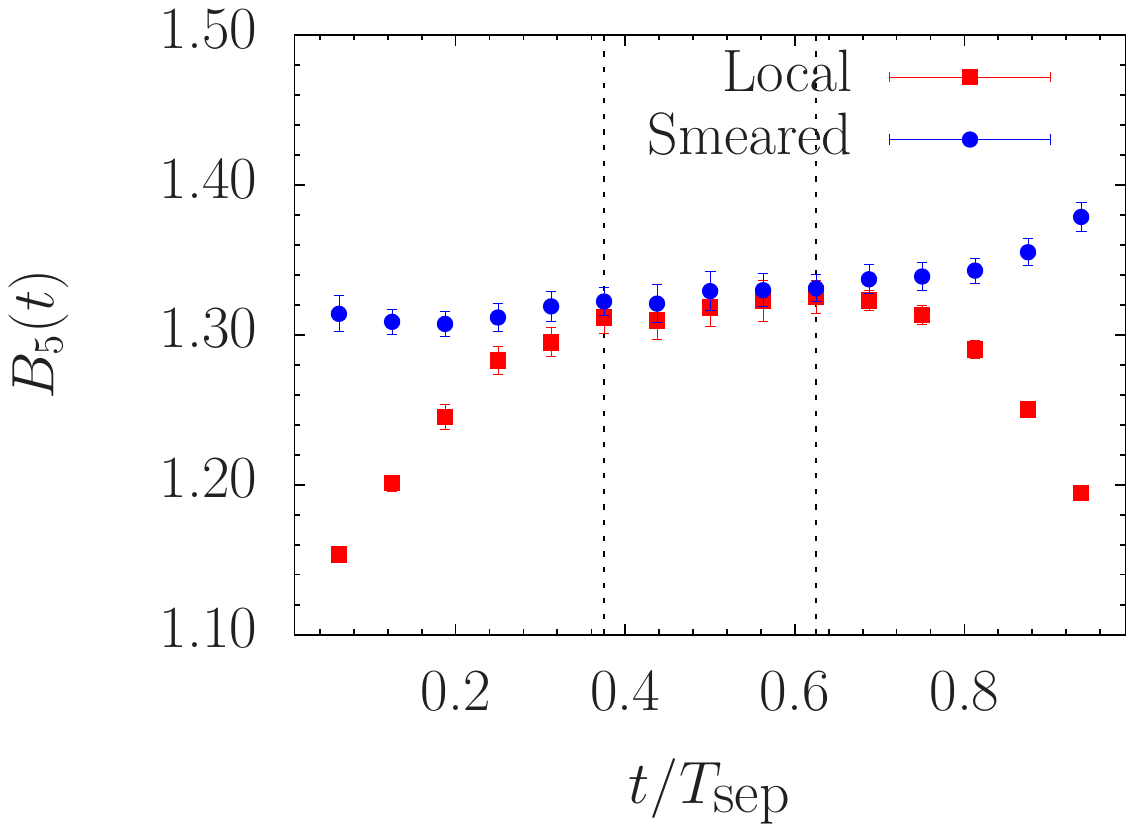}\tabularnewline
\end{tabular}
\end{center}
\vspace*{-0.6cm}
\caption{\label{fig:sme-test} {\it $B_1(t)$ (left) and $B_5(t)$ (right) using either smeared or local sources at $\beta=3.80$ and $(a\mu_{\ell},a\mu_{c})=(0.0080,0.2331)$ on a $24^3\times48$ lattice. The dotted vertical lines delimit the plateau regions.}}
\end{figure}

In fig.~\ref{fig:sme-test} we show (time-dependent estimators of) the $B_1$ and $B_5$ bare bag-parameters at $\beta=3.80$ for the smallest light quark mass and a charm-like quark around the physical charm, and compare the cases of smeared versus local quark sources in both the heavy-light meson interpolating fields.

The bare bag-parameters can be evaluated from ratios of 3-point, 
$C_{3;i}(x_0)$, and two 2-point, $C_2(x_0)$ and $C_2^{\prime}(x_0)$, correlation functions 
(for more details see the discussion that leads to Eqs.~(4.10)-(4.13) of Ref.~\cite{Bertone:2012cu}):
\begin{equation} 
\xi_i\,B_{i}(x_0) \,= \, \dfrac{C_{3;i}(x_0)}{C_{2}(x_0) \,\, C_{2}^\prime(x_0)}, \quad  i=1, \ldots, 5\,.
\label{BBratio}
\end{equation}
To improve the signal-to-noise ratio a sum was performed over the spatial position of the four-fermion operator in $C_{3;i}(x_0)$, and for each gauge configuration the time slice $y_0$ was randomly chosen. 
An important reduction of statistical fluctuations comes also from summing over the spatial position
of both (local or smeared) meson interpolating fields. These spatial sums were implemented at a reasonably low computational cost by means of the stochastic technique discussed in sect. 2.2 of ref.~\cite{Constantinou:2010qv}.

The plateau of the ratio~(\ref{BBratio}), for large source time separation $T_{sep}$, provides an estimate of the (bare) $B_i$ ($i=1, \ldots, 5$) bag-parameter multiplied by the corresponding factor $\xi_i$ in eq.~(\ref{eq:xi}). 
By employing smeared interpolating operators for the meson sources we are able to reduce the source time separation,  $T_{sep}$. The latter, in order to lead to safe plateau signals, turns out to be less than half of the lattice time extension: $T_{sep}/a=\{16, 18, 22, 28\}$ for $\beta=\{3.80, 3.90, 4.05, 4.22\}$, respectively.

\begin{figure}[!t]
\begin{center}
\begin{tabular}{cc}
\hspace*{0.3cm}\vspace*{0.35cm}\includegraphics[bb=171bp 497bp 496bp 705bp,scale=0.6]{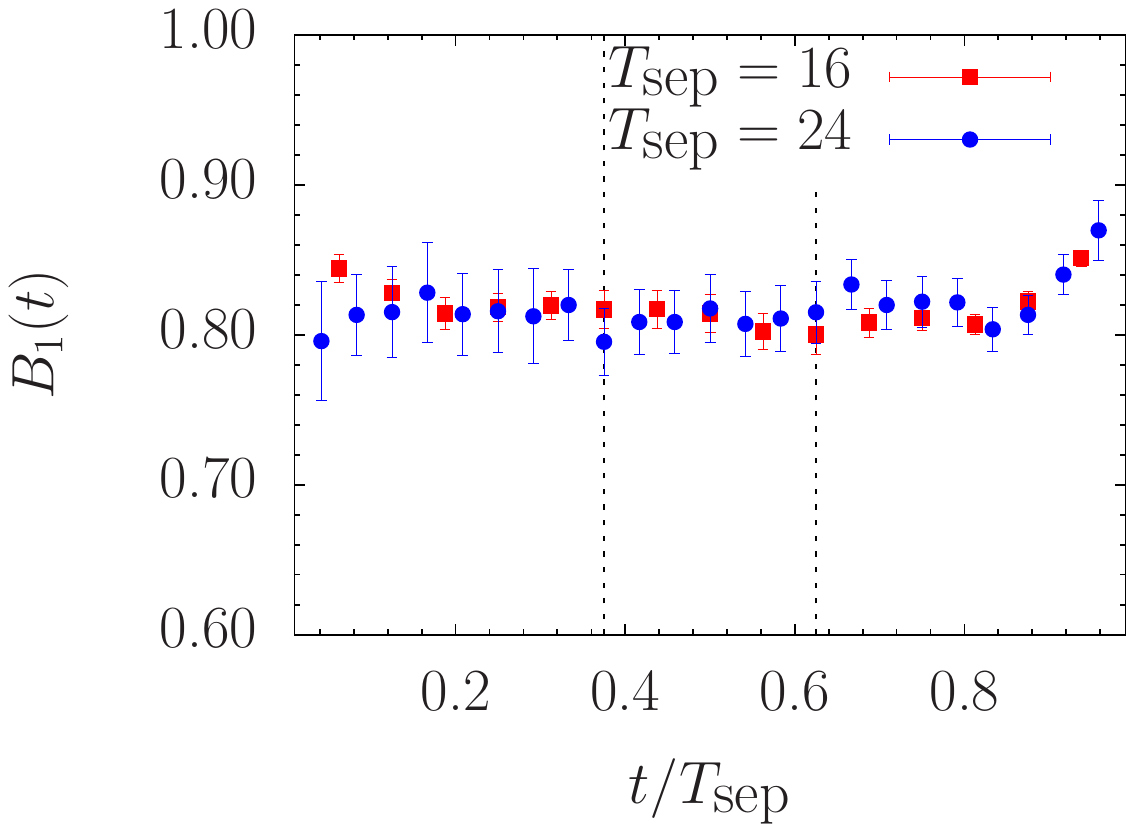} & \vspace*{0.35cm}\includegraphics[bb=171bp 497bp 496bp 705bp,scale=0.6]{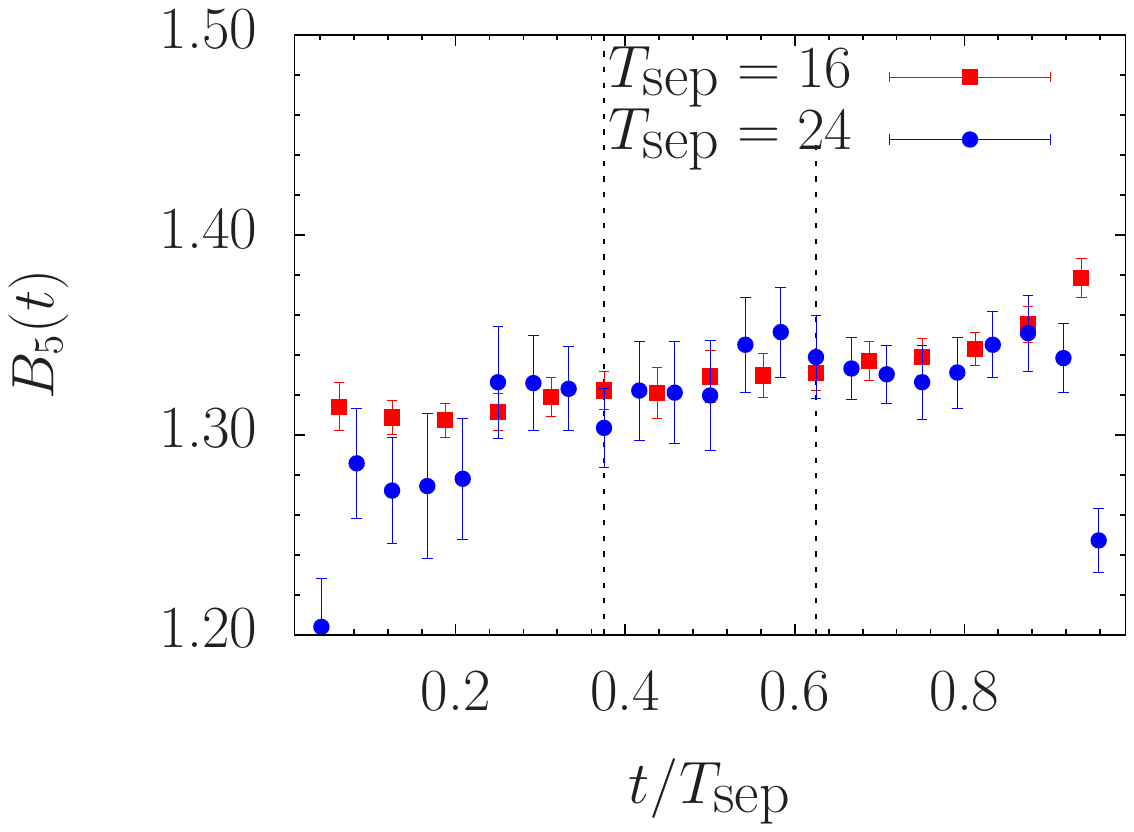}\tabularnewline
\end{tabular}
\end{center}
\vspace*{-0.6cm}
\caption{\label{fig:Tsep-test} $B_1(t)$ (left) and $B_5(t)$ (right) at $\beta=3.80$ and $(a\mu_{\ell},a\mu_{c})=(0.0080,0.2331)$ on a $24^3\times48$ lattice for smeared sources and sink located at two different time distances ($T_{\rm{sep}}$). }
\end{figure}

For illustration, in fig.~\ref{fig:Tsep-test} we show an exploratory test of the effect of locating the source and sink fields at different time slices.
We observe that for both choices, $T_{sep}=16$ and $T_{sep}=24$, there is a visible plateau. Choosing $T_{sep}=16$, as in the present analysis, one obtains data that are more precise than in the $T_{sep}=24$ case.

\section{Results for the bag-parameters at the physical point}
\label{sec:results}
The bag parameters are renormalized non-perturbatively by using the RI-MOM~\cite{RIMOM} renormalization constants computed in ref.~\cite{Bertone:2012cu}.

For all bag-parameters $B_{i}$ the results are first interpolated
to the physical value of the charm quark mass~\cite{Blossier:2010cr}. Since we have simulated
three points around the physical charm quark mass, the interpolation is under
very good control and a linear interpolation turns out to describe
correctly the smooth mass dependence. 

Continuum and chiral extrapolation are carried out in a combined way. For all bag-parameters, we have
tried out a linear fit in the light quark mass, $\bar \mu_\ell$, renormalized in $\overline{\rm{MS}}$ at 2 GeV,
\begin{equation}
B_{i}=A_i\, +B_i\, \bar\mu_{\ell}+D_i\,a^{2}\,, \label{eq:linfit}
\end{equation}
a quadratic fit
\begin{equation}
B_{i}=A'_i+B'_i\,\bar\mu_{\ell}+C'_i\,\bar\mu_{\ell}^{2}+D'_i\,a^{2}\,,\label{eq:quadfit}
\end{equation} and a Heavy Meson Chiral Perturbation Theory (HMChPT) fit ansatz~\cite{Becirevic:2006me}
\begin{equation}
\begin{array}{l}
B_{1}=B_{1}^{\chi}\left[1+b_1\bar{\mu}_{\ell}-\dfrac{(1-3\hat{g}^{2})}{2}\dfrac{2B_{0}\bar\mu_{\ell}}{16\pi^{2}f_{0}^{2}}\log\dfrac{2B_{0}\bar\mu_{\ell}}{16
\pi^{2}f_{0}^{2}}\right]+\hat{D}_1 a^{2}\,,\\
\\
B_{i}=B_{i}^{\chi}\left[1+b_i\bar{\mu}_{\ell}\mp\dfrac{(1\mp 3\hat{g}^{2}Y)}{2}\dfrac{2B_{0}\bar\mu_{\ell}}{16\pi^{2}f_{0}^{2}}\log\dfrac{2B_{0}\bar\mu_{\ell}}
{16\pi^{2}f_{0}^{2}}\right]+\hat{D}_i a^{2}\,,
\end{array}
\end{equation}
where the sign in front of the logarithmic term is minus for $i=2$ and plus for $i=4,5$. We take the HMChPT based estimate $Y=1$ from ref.~\cite{Becirevic:2006me}
and $\hat{g}=0.53(4)$ from the ($N_{f}=2$) lattice measurement of the $g_{D^{*}D\pi}$ coupling~\cite{Becirevic:2012pf}.
We observe that the contribution of the $\hat{g}$ uncertainty to the error of the chiral fit is less than 0.3$\%$ and that of the uncertainties due to
$B_0$ and $f_0$ is less than 0.1$\%$.
In HQET the bag-parameter $B_3$ is related to the bag-parameters $B_1$ and $B_2$. For $Y=1$, which is the only case considered in this paper, the chiral expansion for $B_3$ is similar to the one of $B_2$ with the same chiral log.

\begin{figure}[!ht]
\begin{center}
\begin{tabular}{cc}
\vspace*{0.5cm}\includegraphics[bb=171bp 497bp 496bp 705bp,scale=0.55]{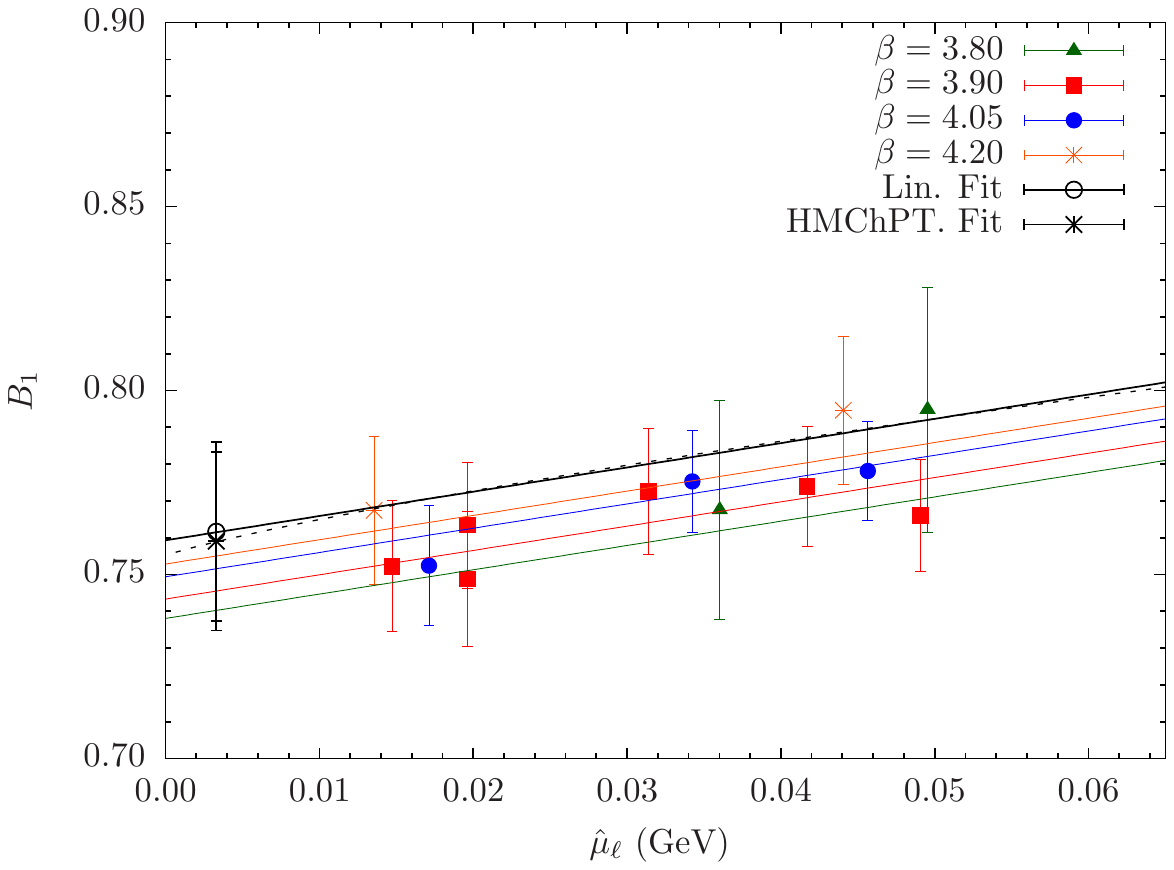} & \vspace*{0.5cm}\includegraphics[bb=171bp 497bp 496bp 705bp,scale=0.55]{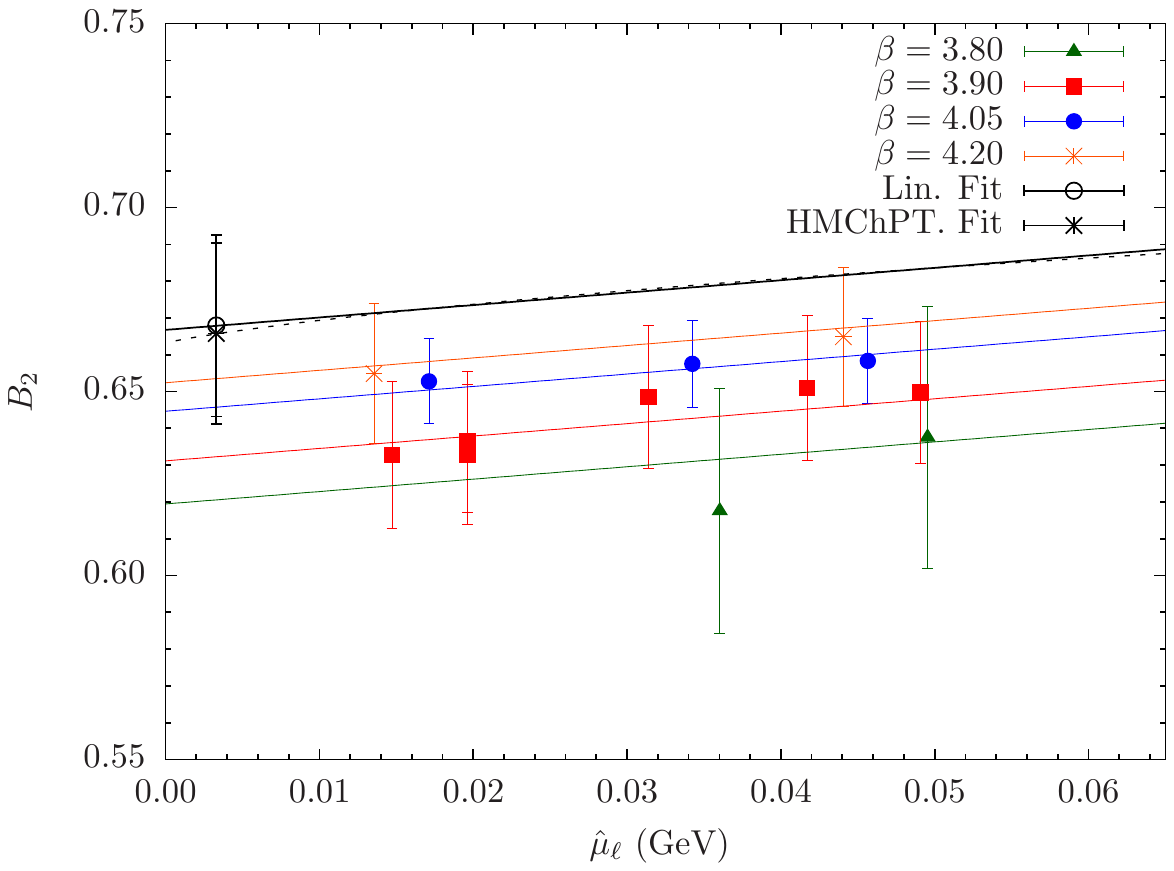}\tabularnewline
\vspace*{0.5cm}\includegraphics[bb=171bp 497bp 496bp 705bp,scale=0.55]{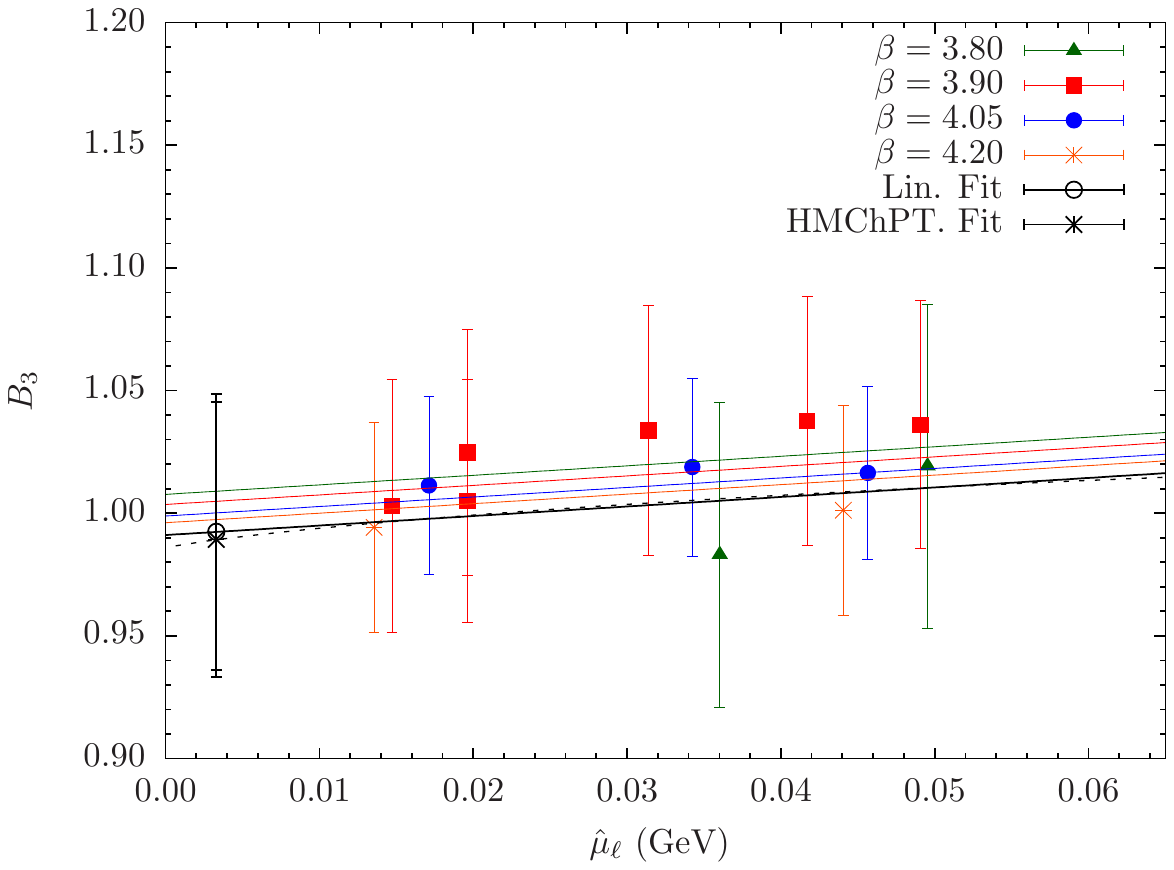} & \vspace*{0.5cm}\includegraphics[bb=171bp 497bp 496bp 705bp,scale=0.55]{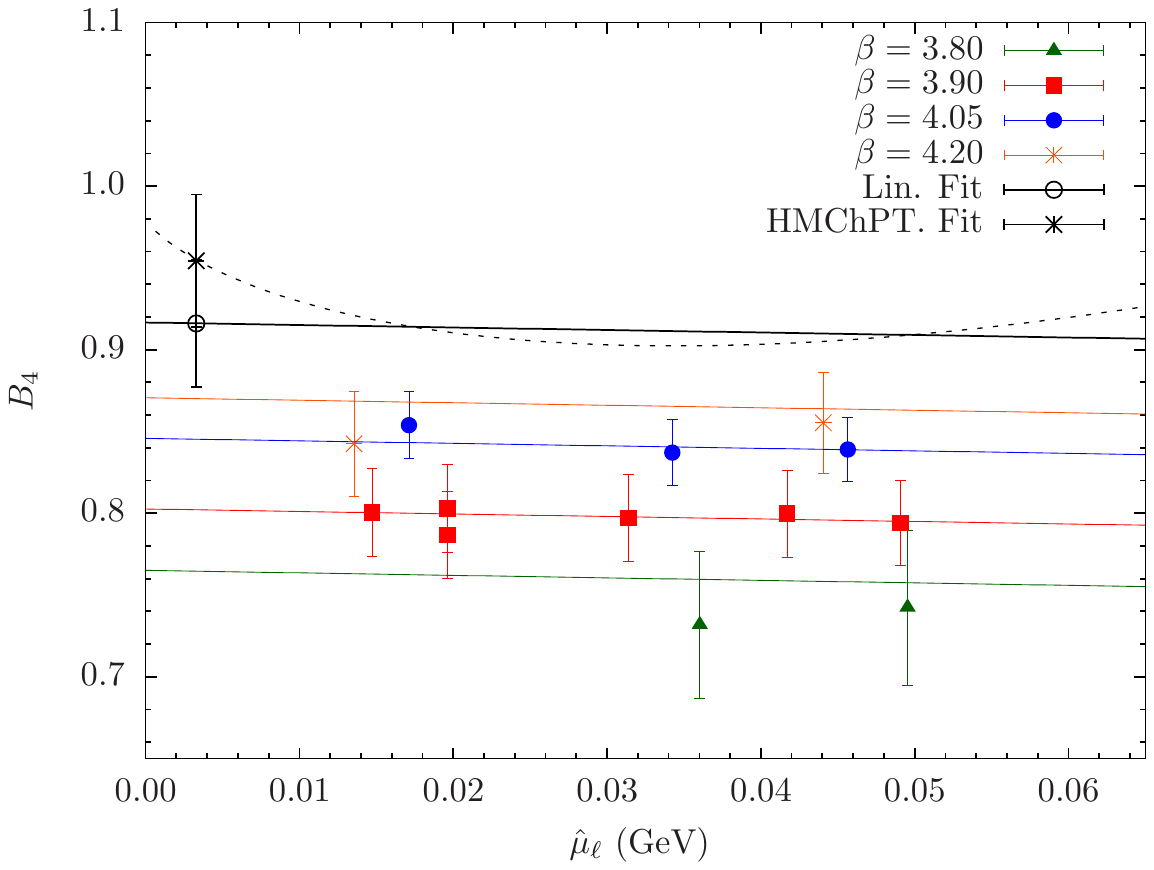}\tabularnewline
\vspace*{0.5cm}\includegraphics[bb=171bp 497bp 496bp 705bp,scale=0.55]{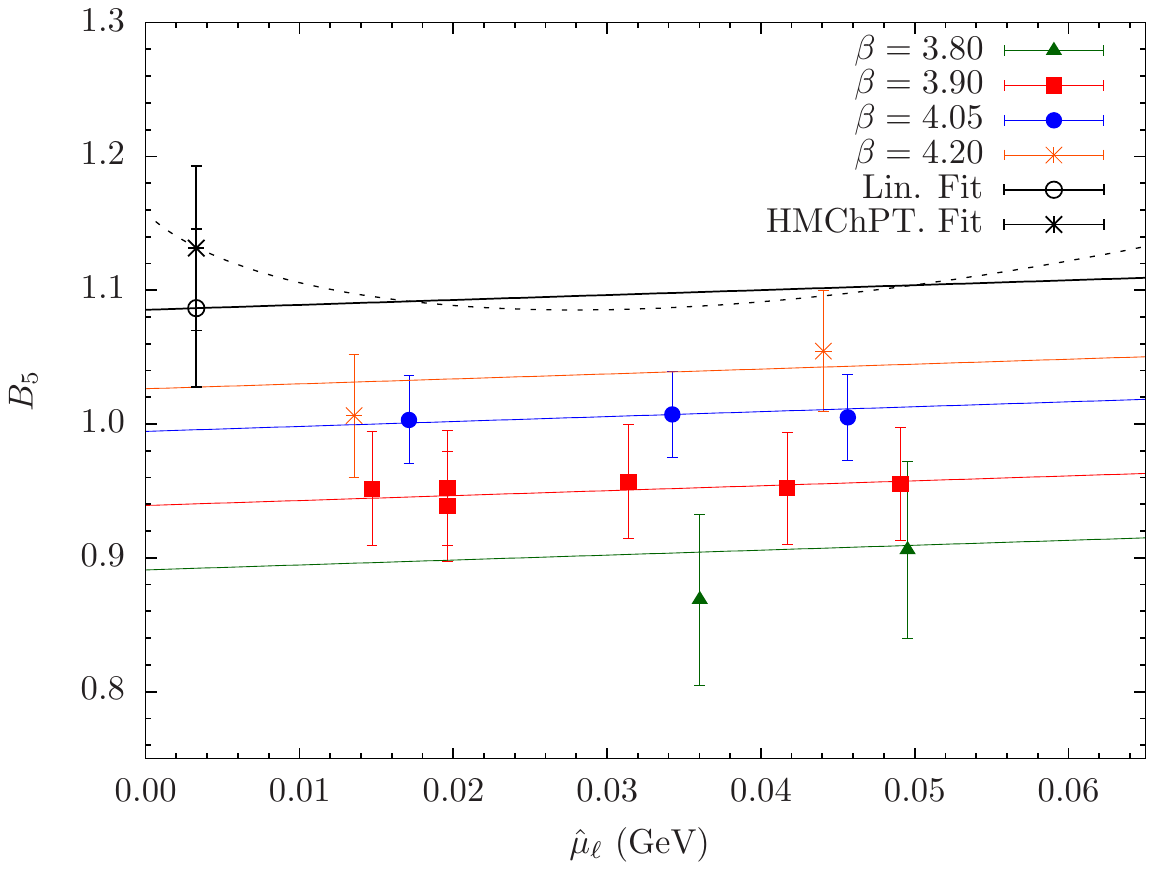} & \vspace*{0.5cm}\tabularnewline
\end{tabular}
\end{center}
\vspace*{-0.6cm}
\caption{\label{fig:BD-continuumlimit}{\it Combined chiral and continuum
extrapolation for the $B_{i}$ parameters ($i=1,2,3,4,5$) renormalized in $\overline{\rm{MS}}$
scheme of ref.~\cite{mu:4ferm-nlo} at 3 GeV.
Solid lines represent the linear chiral fit with the continuum curve displayed in black.
The dashed black line represents the continuum curve in the case of the HMChPT ansatz. Open
circles and stars stand for the results at the physical point
corresponding to the linear and HMChPT fit respectively.}}
\end{figure}

In fig.~\ref{fig:BD-continuumlimit} we show the combined chiral and continuum fit for the renormalized $B_{i}$ in the $\overline{\rm{MS}}$ scheme of
ref.~\cite{mu:4ferm-nlo} at 3 GeV.
Our final results for the $B_{i}$ bag-parameters in the $\overline{\rm{MS}}$ and
RI-MOM scheme at 3 GeV, obtained by averaging the estimates from the three chiral fits discussed above, are collected in Table~\ref{tab:BD-results}. The quadratic fit results turn out to be very close to those of the linear fit. The half of the difference between the two more distant results, i.e. between the results of the linear and HMChPT fits, has been included as a systematic error, added in quadrature to the statistical one.
In performing the combined chiral and continuum fits statistical errors of the (bare) bag parameters 
and statistical uncertainties of the renormalisation constants of two- and four-fermion operators 
have been included and treated altogether employing the bootstrap procedure.  
We note that statistical errors of the renormalisation constants represent a significant source of uncertainty. 
Their contibution in the final error budget lies between 2$\%$ and 3.5$\%$, depending on $B_i$. 
The largest one is noted for $B_3$. 
Moreover, we have added in quadrature the systematic error owed to the way that discretisation effects have been estimated 
in computing 
the renormalisation constants\footnote{More details on the RI-MOM computation of the renormalisation constants 
and the two possible ways to work out estimates concerning discretisation errors can be found in refs.~\cite{Bertone:2012cu,Constantinou:2010gr}.}.
This systematic uncertainty varies from 0.5$\%$ to 2.5$\%$.

\vspace*{1.cm}
\noindent{\bf Acknowledgements}
We acknowledge the computer time available on the Altix system at the HLRN supercomputing service in Berlin
under the project "B-physics from lattice QCD simulations". Part of this work has been completed thanks to allocation of  CPU time on BlueGene/Q -Fermi based on the
agreement between INFN and CINECA within the specific initiative INFN-RM123.
The research leading to these results has received funding from the European Research Council under the European Union's Seventh Framework Programme (FP/2007-2013) / ERC Grant Agreements n. 279972 ``NPFlavour'' and n. 267985 ``DaMeSyFla''.
F.S. has received funding from the
European Research Council under the European Union's Seventh
Framework Programme (FP7/2007-2013) ERC grant agreement n. 279757.
We also thank MIUR (Italy) for partial support under the contract PRIN10-11.

\appendix
\section{\label{sec:table_results}Lattice data for the bare bag parameters}

In this appendix we collect the results for the {\it bare} bag parameters, for all simulated values of $\beta$
and combinations of quark masses.


\begin{table}[H]
\begin{center}\begin{tabular}{|cc|ccccc|} 
\hline
\multicolumn{7}{|c|} {$\beta=3.80$  $L^3 \times T = 24^3 \times 48 $ }\\
\hline
$a\mu_{\ell}=a\mu_{sea}$ & $a\mu_c$ &  $\xi_1 B_1$ & $\xi_2 B_2$ & $\xi_3 B_3$ & $\xi_4
 B_4$ & $\xi_5 B_5$  \\
\hline
0.0080 & 0.1982 & 2.126(30) & 1.285(09) &  0.289(03) &  2.120(10) &  0.860(06) 
  \\
 & 0.2331  & 2.160(32) & 1.307(09) &  0.291(03) &  2.135(11) &  0.884(07)  \\
 & 0.2742  & 2.193(35) & 1.329(10) &  0.291(03) &  2.148(12) &  0.909(08)  \\
\hline
0.0110 & 0.1982 & 2.196(43) & 1.326(24) &  0.299(05) &  2.153(35) &  0.884(15) 
  \\
 & 0.2331  & 2.236(44) & 1.349(24) &  0.301(05) &  2.168(36) &  0.910(15)  \\
 & 0.2742  & 2.277(45) & 1.373(24) &  0.303(05) &  2.181(37) &  0.936(16)  \\
\hline
\end{tabular}\end{center}
\caption{Bare $\xi_{i}B_{i}$  at each combination
of the quark mass pair $(a\mu_{l},a\mu_{c})$ at $\beta=3.80$ and
$24^{3}\times48$ volume. }
\label{tab:b380}
\end{table}
\begin{table}[H]
\begin{center}\begin{tabular}{|cc|ccccc|} 
\hline
\multicolumn{7}{|c|} {$\beta=3.90$  $L^3 \times T = 24^3 \times 48 $ }\\
\hline
$a\mu_{\ell}=a\mu_{sea}$ & $a\mu_c$ &  $\xi_1 B_1$ & $\xi_2 B_2$ & $\xi_3 B_3$ & $\xi_4
 B_4$ & $\xi_5 B_5$  \\
\hline
0.0040 & 0.1828 & 2.089(28) & 1.245(07) &  0.284(02) &  2.169(14) &  0.883(08) 
  \\
 & 0.2150  & 2.125(31) & 1.269(07) &  0.287(03) &  2.186(15) &  0.910(08)  \\
 & 0.2529  & 2.157(34) & 1.292(08) &  0.289(03) &  2.202(16) &  0.938(10)  \\
\hline
0.0064 & 0.1828 & 2.152(24) & 1.269(08) &  0.288(02) &  2.159(16) &  0.884(07) 
  \\
 & 0.2150  & 2.192(26) & 1.293(08) &  0.290(02) &  2.171(16) &  0.910(07)  \\
 & 0.2529  & 2.231(28) & 1.317(08) &  0.291(03) &  2.182(16) &  0.936(07)  \\
\hline
0.0085 & 0.1828 & 2.155(18) & 1.274(06) &  0.289(01) &  2.160(14) &  0.881(07) 
  \\
 & 0.2150  & 2.196(20) & 1.298(07) &  0.291(02) &  2.178(15) &  0.908(07)  \\
 & 0.2529  & 2.235(21) & 1.322(07) &  0.293(02) &  2.194(16) &  0.937(08)  \\
\hline
0.0100 & 0.1828 & 2.136(12) & 1.273(05) &  0.288(02) &  2.148(11) &  0.880(05) 
  \\
 & 0.2150  & 2.173(12) & 1.295(06) &  0.291(02) &  2.163(12) &  0.907(06)  \\
 & 0.2529  & 2.209(13) & 1.318(06) &  0.292(02) &  2.176(13) &  0.934(06)  \\
\hline
\end{tabular}\end{center}
\caption{The same as in Table \ref{tab:b380} but for $\beta=3.90$ and
$24^{3}\times48$ volume.}
\end{table}
\begin{table}[H]
\begin{center}\begin{tabular}{|cc|ccccc|} 
\hline
\multicolumn{7}{|c|} {$\beta=3.90$  $L^3 \times T = 32^3 \times 64 $ }\\
\hline
$a\mu_{\ell}=a\mu_{sea}$ & $a\mu_c$ &  $\xi_1 B_1$ & $\xi_2 B_2$ & $\xi_3 B_3$ & $\xi_4
 B_4$ & $\xi_5 B_5$  \\
\hline
0.0030 & 0.1828 & 2.099(28) & 1.238(12) &  0.279(04) &  2.165(16) &  0.881(07) 
  \\
 & 0.2150  & 2.134(30) & 1.262(13) &  0.282(04) &  2.179(17) &  0.908(08)  \\
 & 0.2529  & 2.168(32) & 1.286(14) &  0.284(05) &  2.193(19) &  0.937(09)  \\
\hline
0.0040 & 0.1828 & 2.119(25) & 1.239(06) &  0.281(02) &  2.128(16) &  0.868(05) 
  \\
 & 0.2150  & 2.165(26) & 1.262(07) &  0.282(02) &  2.142(17) &  0.895(05)  \\
 & 0.2529  & 2.212(27) & 1.285(07) &  0.283(02) &  2.155(18) &  0.923(06)  \\
\hline
\end{tabular}\end{center}
\caption{The same as in Table \ref{tab:b380} but for $\beta=3.90$ and
$32^{3}\times64$ volume.}
\end{table}
\begin{table}[H]
\begin{center}\begin{tabular}{|cc|ccccc|} 
\hline
\multicolumn{7}{|c|} {$\beta=4.05$  $L^3 \times T = 32^3 \times 64 $ }\\
\hline
$a\mu_{\ell}=a\mu_{sea}$ & $a\mu_c$ &  $\xi_1 B_1$ & $\xi_2 B_2$ & $\xi_3 B_3$ & $\xi_4
 B_4$ & $\xi_5 B_5$  \\
\hline
0.0030 & 0.1572 & 2.058(29) & 1.213(08) &  0.277(03) &  2.183(16) &  0.881(07) 
  \\
 & 0.1849  & 2.088(32) & 1.234(10) &  0.279(03) &  2.202(17) &  0.910(07)  \\
 & 0.2175  & 2.114(36) & 1.255(12) &  0.280(04) &  2.219(19) &  0.940(08)  \\
\hline
0.0060 & 0.1572 & 2.115(23) & 1.221(08) &  0.279(03) &  2.144(15) &  0.875(07) 
  \\
 & 0.1849  & 2.151(25) & 1.243(09) &  0.281(03) &  2.160(16) &  0.904(08)  \\
 & 0.2175  & 2.185(28) & 1.265(10) &  0.282(04) &  2.175(18) &  0.933(09)  \\
\hline
0.0080 & 0.1572 & 2.121(20) & 1.223(08) &  0.279(02) &  2.149(11) &  0.877(05) 
  \\
 & 0.1849  & 2.158(22) & 1.245(08) &  0.280(02) &  2.165(12) &  0.904(06)  \\
 & 0.2175  & 2.195(25) & 1.266(09) &  0.281(03) &  2.179(12) &  0.931(07)  \\
\hline
\end{tabular}\end{center}
\caption{The same as in Table \ref{tab:b380} but for $\beta=4.05$ and
$32^{3}\times64$ volume.}
\end{table}
\begin{table}[H]
\begin{center}\begin{tabular}{|cc|ccccc|} 
\hline
\multicolumn{7}{|c|} {$\beta=4.20$  $L^3 \times T = 32^3 \times 64 $ }\\
\hline
$a\mu_{\ell}=a\mu_{sea}$ & $a\mu_c$ &  $\xi_1 B_1$ & $\xi_2 B_2$ & $\xi_3 B_3$ & $\xi_4
 B_4$ & $\xi_5 B_5$  \\
\hline
0.0065 & 0.13315 & 2.128(32) & 1.205(12) &  0.278(04) &  2.157(20) &  0.878(10)
   \\
 & 0.1566  & 2.174(33) & 1.229(13) &  0.280(05) &  2.178(20) &  0.908(11)  \\
 & 0.1842  & 2.220(33) & 1.253(15) &  0.282(05) &  2.199(21) &  0.940(12)  \\
\hline
\end{tabular}\end{center}
\caption{The same as in Table \ref{tab:b380} but for $\beta=4.20$ and
$32^{3}\times64$ volume.}
\end{table}
\begin{table}[H]
\begin{center}\begin{tabular}{|cc|ccccc|} 
\hline
\multicolumn{7}{|c|} {$\beta=4.20$  $L^3 \times T = 48^3 \times 96 $ }\\
\hline
$a\mu_{\ell}=a\mu_{sea}$ & $a\mu_c$ &  $\xi_1 B_1$ & $\xi_2 B_2$ & $\xi_3 B_3$ & $\xi_4
 B_4$ & $\xi_5 B_5$  \\
\hline
0.0020 & 0.13315 & 2.068(33) & 1.187(12) &  0.275(04) &  2.130(28) &  0.853(12)
   \\
 & 0.1566  & 2.100(33) & 1.211(13) &  0.278(05) &  2.143(29) &  0.879(13)  \\
 & 0.1842  & 2.131(34) & 1.234(15) &  0.280(06) &  2.156(31) &  0.905(14)  \\
\hline
\end{tabular}\end{center}
\caption{The same as in Table \ref{tab:b380} but for $\beta=4.20$ and
$48^{3}\times96$ volume.}
\end{table}

\clearpage

\bibliographystyle{mybibstyle}
\bibliography{lattice}

\end{document}